\documentclass[aps,prb,reprint,amsmath,amssymb,superscriptaddress]{revtex4-2}
\usepackage{graphicx}
\usepackage{mathtools}
\usepackage{bm}
\usepackage[dvipsnames]{xcolor}
\usepackage{hyperref}
\hypersetup{
    colorlinks  = true,
    linkcolor   = BrickRed,
    citecolor   = MidnightBlue,
    urlcolor    = MidnightBlue
}
\usepackage{orcidlink}

\newcommand{\ve}[1]{\mathbf{#1}}
\newcommand{\n}{\ve{n}} 
\renewcommand{\r}{\ve{r}} 
\newcommand{\E}{\ve{E}} 
\newcommand{\del}{\partial} 
\newcommand{\grad}{\bm{\nabla}} 

\begin{document}
\title{The surprising subtlety of electrostatic field lines}

\author{Kevin Zhou\,\orcidlink{0000-0002-9810-3977}}
\email{kzhou7@berkeley.edu}
\affiliation{Berkeley Center for Theoretical Physics, University of California, Berkeley, CA 94720, USA}

\author{Tom\'{a}\v{s} Brauner\,\orcidlink{0000-0002-5564-4845}}
\email{tomas.brauner@uis.no}
\affiliation{Department of Mathematics and Physics, University of Stavanger, 4036 Stavanger, Norway}

\begin{abstract}
Electric fields are commonly visualized with field line diagrams, which only unambiguously specify the field's direction. We consider two simple questions. First, can one deduce if an electric field is conservative, as required e.g.~in electrostatics, from its field lines alone? Second, are there conservative electric fields with straight field lines, besides the familiar textbook examples with spherical, cylindrical, or planar symmetry? We give a self-contained introduction to the differential geometry required to answer these questions, assuming only vector calculus background. 
\end{abstract}

\maketitle


\section{Introduction}
\label{sec:intro}

In electromagnetism, the electric and magnetic fields are the fundamental quantities, and field lines give only incomplete information about the fields. However, field lines remain a useful source of visual intuition. For example, the directions of electric and magnetic forces can be seen from field line diagrams~\cite{10.1119/1.15925}, using Faraday's idea that field lines carry a tension along them and an outward pressure perpendicular to them. Radiation from a kicked charge is readily visualized in terms of kinked electric field lines~\cite{Purcell2013}, and magnetic field lines in plasmas can be thought of as ``moving'' with the medium~\cite{newcomb1958motion,stern1966motion,10.1119/1.1531577}. 

Field line diagrams are almost universally used in introductory physics courses, and it is well-known that they can lead to common student misconceptions~\cite{10.1119/1.17265,pocovi2002lines,10.1119/1.5100588,PhysRevPhysEducRes.12.020134}. However, there are also legitimate questions that students may ask about them, which have surprisingly subtle answers. In this work, we consider two such questions.

First, electric field lines are often used to depict electrostatic fields, which are conservative. But given a set of field lines, how can one tell if they can correspond to a conservative field? Second, introductory courses usually illustrate Gauss's law by applying it to electrostatic fields with spherical, cylindrical, or planar symmetry, as these simple cases have only straight field lines. Are there any other electrostatic fields with only straight field lines? 

As we will discuss in Sec.~\ref{sec:graphrep}, these questions are nontrivial because the field's magnitude is not determined by the field lines. We answer the first question in Sec.~\ref{sec:integrability}. It turns out to require the geometric notion of Frobenius integrability, which is not commonly taught in undergraduate physics courses, but can be introduced with only vector calculus background. The second question is most simply answered by considering the Gaussian and mean curvatures of equipotential surfaces. We give a self-contained introduction to these concepts in Sec.~\ref{sec:curvature}, answer the question in Sec.~\ref{sec:straightlines}, and conclude in Sec.~\ref{sec:discussion}.

These puzzles may be of interest to instructors, and their resolutions may expand the horizons of ambitious upper-division students. For more about the underlying geometrical ideas, we direct the reader to Refs.~\cite{schutz1980geometrical,Frankel2012a,do2016differential,needham2021visual}.


\section{Visualizing Electric Fields}
\label{sec:graphrep}

\begin{figure*}
\includegraphics[width=0.7\textwidth]{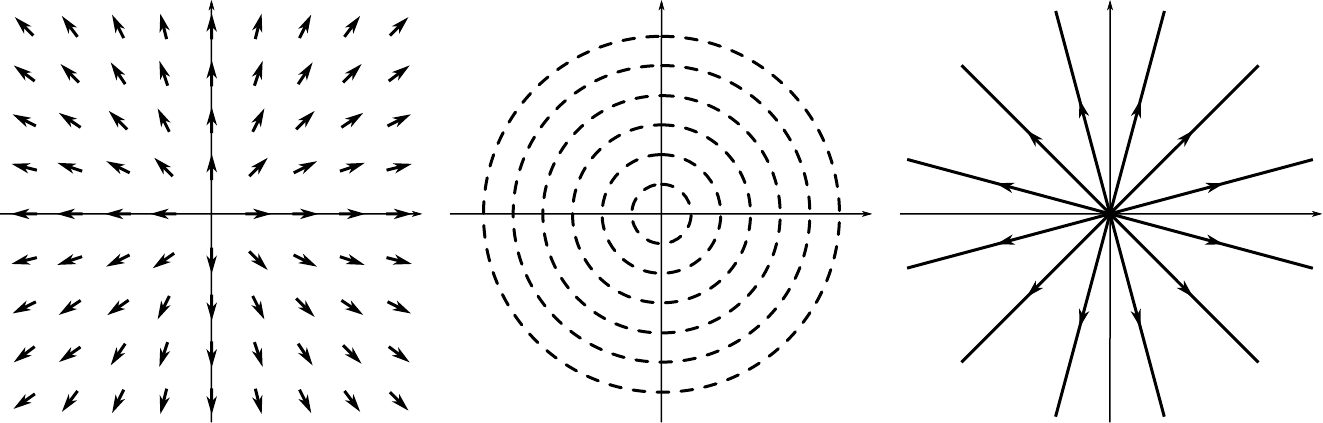}
\caption{A vector field can be represented graphically by showing its field vector at a set of points, a set of equipotential surfaces (if the field is conservative), or a set of field lines (i.e.~integral curves, always tangent to the field).}
\label{fig:fieldvisualization}
\end{figure*}

Electric fields can be visualized by drawing:
\begin{itemize}
\item[(a)] the field vector at a grid of points in space,
\item[(b)] a set of \textit{unlabeled} equipotential surfaces, or
\item[(c)] a set of oriented, \textit{maximally extended} field lines.
\end{itemize}
These three common methods, illustrated in Fig.~\ref{fig:fieldvisualization}, encode different partial information about the field. 

Field vector plots show both the field's magnitude and direction, and allow two fields to be naturally superposed by vector addition at each point. However, such diagrams can be visually cluttered, and are too time-consuming to produce by hand in a lecture setting. 

Equipotential surfaces only make sense for conservative fields, and do not show the direction of the field as directly as the other methods. In addition, they do not fully determine the magnitude of the field, since multiplying the field by any scalar function constant on the equipotential surfaces keeps them the same. For example, any electric field of the form $\E(\r) = (0,0,f(z))$ has planar equipotential surfaces parallel to the $xy$-plane. 

Field lines efficiently show the field's direction everywhere. They are the quickest to draw, but also the most ambiguous, since multiplying a field by any positive scalar function $f(\r)$ keeps the field lines the same. For instance, all electric fields of the form $\E(\r) = (0,0,f(\r))$ have field lines parallel to the $z$-axis. Field lines are ill-defined at points where the field is zero or infinite, but we will ignore such special points, as our discussion will only deal with the field's local structure.

It is possible to draw a finite number of field lines so that their local density roughly indicates the field's magnitude. However, this requires a special prescription for beginning and ending field lines, and is generally not possible in two-dimensional diagrams~\cite{10.1119/1.17939,10.1119/1.18237}. Instead, we will consider the set of \textit{all} field lines, each extended as far as possible. This is equivalent to specifying a unit vector field $\n(\r)$ which gives the direction of the electric field everywhere in space. 


\section{Field Lines Corresponding to Electrostatic Fields}
\label{sec:integrability}

First, we consider when a given set of field lines can correspond to an electrostatic field. That is, given $\n(\r)$, when is it possible to define a nonzero $f(\r)$ so that $\E = f \n$ is conservative? In this section we will show that this is possible, at least locally, if and only if $\n$ satisfies
\begin{equation} \label{eq:integratingfactor}
\n\cdot(\grad\times\n)=0. 
\end{equation}

It is easy to show this condition is necessary. Locally, a field is conservative precisely when it is curl-free, but
\begin{equation}
\grad\times\E=(\grad f)\times\n + f \, \grad\times\n.
\end{equation}
The first term is perpendicular to $\n$, so it can only cancel the second term if $\grad \times \n$ is also perpendicular to $\n$. 

However, it is harder to show that Eq.~\eqref{eq:integratingfactor} is sufficient. If it holds at a point, we can certainly choose $f$ and $\grad f$ so that $\grad \times \E$ vanishes at that point. But if it holds in an entire region, it is not clear if $f$ can be consistently defined throughout that region. Proceeding along these lines turns out to be challenging, and it is easier to forget about $f$ and instead think about equipotential surfaces.

If $\E = f \n$ is conservative, it has a set of equipotential surfaces, which are always perpendicular to it, and thus also perpendicular to $\n$. Conversely, specifying a general $\n$ defines an infinitesimal plane at each point. This $\n$ can correspond to a conservative $\E$ precisely when those planes ``mesh together'' into distinct equipotential surfaces. In this case, we say that $\n$ is integrable.

\begin{figure}
\includegraphics[width=0.4\columnwidth]{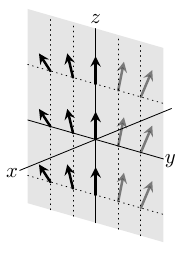} \includegraphics[width=0.4\columnwidth]{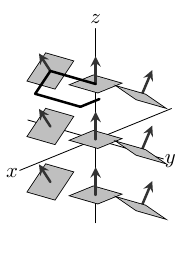}
\caption{An $\n(\r)$ which is not integrable. Left: a field vector plot of $\n(\r)$ in the $x = 0$ plane. This vector field is independent of $x$ and $z$, and rotates about the $y$-axis as one moves along it. Right: if one follows the planes orthogonal to $\n$ in a rectangle above the $xy$ plane, one returns shifted along the $z$-axis. Thus, the planes cannot mesh together into equipotential surfaces.}
\label{fig:planes}
\end{figure}

Figure~\ref{fig:planes} shows an $\n$ proportional to $(-y,0,1)$, which violates Eq.~\eqref{eq:integratingfactor} and is therefore not integrable. To see this geometrically, consider moving from a point on the $z$-axis along the $y$-axis, then turning by $90^\circ$ along the planes, and repeating this process three times. Instead of arriving back at the same point, one returns shifted in the $z$-direction. By repeating this process, one can reach any point on the $z$-axis, and from there any point in space. Yet, all of these points should be part of the same putative equipotential surface. This is a contradiction, so we conclude that it is impossible to slice the space into distinct equipotential surfaces of $\n$.

Suppose now that Eq.~\eqref{eq:integratingfactor} holds, and that $\grad \times \n$ is nonzero, as otherwise $\n$ would be trivially integrable. We may then define the vector fields 
\begin{align}
\label{eq:v1def}
\ve{v}_1 &= \grad \times \n, \\
\label{eq:v2def}
\ve{v}_2 &= \n \times \ve{v}_1 / |\ve{v}_1|^2.
\end{align}
These fields satisfy $\ve{v}_1 \times \ve{v}_2 = \n$, so that the vectors $(\ve{v}_1,\ve{v}_2,\n)$ constitute a right-handed orthogonal system. The vectors $\ve{v}_1,\ve{v}_2$ span the tangent plane to the local ``equipotential surface'' of $\n$.

Consider a loop that starts at a point $\ve{r}_0$. Move with velocity $\ve{v}_1$ for an infinitesimal time $\epsilon$, then move with velocity $\ve{v}_2$ for time $\epsilon$, then with $-\ve{v}_1$ and $-\ve{v}_2$, arriving at a final point $\ve{r}_f$. Any finite loop on a putative equipotential surface can be decomposed into such infinitesimal rectangles. Since the number of rectangles in the loop scales as $1/\epsilon^2$, it is sufficient to check if the shift away from the local equipotential surface upon traversing an infinitesimal rectangle vanishes to order $\epsilon^2$. Thus, the field $\n$ is integrable if, for any starting point $\ve{r}_0$, we have
\begin{equation} \label{eq:integrability_cond}
\lim_{\epsilon \to 0} \frac{\n(\ve{r}_0) \cdot (\ve{r}_f - \ve{r}_0)}{\epsilon^2} = 0.
\end{equation}

To evaluate $\ve{r}_f-\ve{r}_0$ to order $\epsilon^2$, it suffices to expand each vector field in a Taylor series about $\ve{r}_0$ to first order,
\begin{equation}
\ve{v}_i(\ve{r}) = \ve{v}_i(\ve{r}_0) + (\ve{r} - \ve{r}_0) \cdot (\grad \ve{v}_i(\ve{r}_0)) + \ldots
\end{equation}
for $i = 1$ or $2$, where $\grad \ve{v}_i$ is the Jacobian matrix, containing derivatives of the components of $\ve{v}_i$. After traversing the first side of the rectangle, we arrive at the point 
\begin{equation}
\ve{r}_1 \simeq \ve{r}_0 + \epsilon \, \ve{v}_1(\ve{r}_0) + \frac{\epsilon^2}{2} \, \ve{v}_1(\ve{r}_0) \cdot (\grad \ve{v}_1(\ve{r}_0))
\end{equation}
where the $\simeq$ indicates that we are dropping terms of third and higher order in $\epsilon$. After the second side, we arrive at 
\begin{equation}
\ve{r}_2 \simeq \ve{r}_1 + \epsilon \, \ve{v}_2(\ve{r}_1) + \frac{\epsilon^2}{2} \, \ve{v}_2(\ve{r}_1) \cdot (\grad \ve{v}_2(\ve{r}_1)).
\end{equation}
In the third term, the argument $\ve{r}_1$ can be replaced with $\ve{r}_0$, since this only produces an order $\epsilon^3$ change. Similarly, we can expand the second term as
\begin{equation}
\epsilon \, \ve{v}_2(\ve{r}_1) \simeq \epsilon \, \ve{v}_2(\ve{r}_0) + \epsilon^2 \, \ve{v}_1(\ve{r}_0) \cdot (\grad \ve{v}_2(\ve{r}_0)).
\end{equation}
We may proceed similarly for the other two sides. All order $\epsilon$ terms in $\ve{r}_f$ cancel, and the final result is 
\begin{equation}
\ve{r}_f - \ve{r}_0 \simeq \epsilon^2 (\ve{v}_1 \cdot (\grad \ve{v}_2) - \ve{v}_2 \cdot (\grad \ve{v}_1))
\end{equation}
where the right-hand side is evaluated at $\ve{r}_0$. 

By generalizing Eq.~\eqref{eq:integrability_cond} to an arbitrary starting point, we thus have shown that $\n$ is integrable if
\begin{equation} \label{eq:frob_int}
\n \cdot ((\ve{v}_1 \cdot \grad) \, \ve{v}_2 - (\ve{v}_2 \cdot \grad) \, \ve{v}_1) = 0.
\end{equation}
To evaluate the left-hand side, we apply the product rule for curl, giving 
\begin{equation}
\n \cdot (\ve{v}_1 (\grad \cdot \ve{v}_2) - \ve{v}_2 (\grad \cdot \ve{v}_1) - \grad \times (\ve{v}_1 \times \ve{v}_2)).
\end{equation}
The first two terms vanish because $\n \cdot \ve{v}_i = 0$, and the final term vanishes by Eq.~\eqref{eq:integratingfactor} because $\ve{v}_1 \times \ve{v}_2 = \ve{n}$. We have therefore shown that Eq.~\eqref{eq:integratingfactor} implies that $\n$ is integrable. 

The above discussion does not produce an explicit expression for the magnitude $f$ of the field. Still, we can gain some information about $f$ indirectly. If Eq.~\eqref{eq:integratingfactor} holds, then the field lines specified by $\n(\r)$ correspond to a set of equipotential surfaces. Suppose we constructed two nearby equipotential surfaces, $S$ and $S'$. Since $\ve{E} = - \grad \phi$, where $\phi$ is the electric potential, the value of $f$ at some point on $S$ is inversely proportional to that point's distance to $S'$. This gives us some information about the variation of $f$ along an equipotential surface. It is however still possible to rescale $f$ by any function constant on the equipotential surfaces. Such rescalings change the field's divergence, i.e.~the physical charge density.

The result Eq.~\eqref{eq:frob_int} is a simple example of the Frobenius integrability criterion. Analogues of it govern when constraints in mechanics are holonomic, and when differentials in thermodynamics correspond to state functions~\cite{schutz1980geometrical,Frankel2012a}. Technically, this criterion only ensures ``local'' integrability: i.e., that $f$ exists in a region near a given point. There can also be ``global'' obstructions to integrability, such as a field line forming a closed loop.


\section{Introducing Curvature}
\label{sec:curvature}

Next, we would like to find the most general electrostatic fields with straight field lines. To proceed, we will need to consider the geometry of the field's equipotential surfaces, which requires first introducing some basic facts about curvature. 

At a given point on a surface, the normal vector $\n$ defines a tangent plane, which serves as a ``first order'' approximation of the surface about that point. Curvature encapsulates the ``second order'' information. 

Concretely, suppose we work in Cartesian coordinates with the origin at some point on the surface, and orient the $z$-axis so that the normal vector there is $\n = (0,0,1)$. Then the surface near that point can be parametrized as 
\begin{equation}
z \simeq c_1 x^2 + c_2 xy + c_3 y^2
\end{equation}
where $\simeq$ denotes equality up to cubic and higher-order terms in $x$ and $y$. We can orient the $x$ and $y$-axes along the ``principal'' directions, where there is no $xy$ term, so
\begin{equation} \label{eq:z_param}
-z \simeq \frac{k_1}{2} \, x^2 + \frac{k_2}{2} \, y^2.
\end{equation}
The sign of this expression is arbitrary, but chosen to reduce the number of minus signs that will appear later. 

Note that to first order in $z$, the intersection of the surface with the $y = 0$ plane is the circle $x^2 + (z + 1/k_1)^2 = 1/k_1^2$. Thus, $R_1 = 1 / k_1$ is the radius of curvature of a curve which follows the surface along the $x$ direction, and similarly $R_2 = 1 / k_2$ is the analogous quantity for the $y$ direction. Since their reciprocals $k_1$ and $k_2$ describe the rates at which these curves turn, they are called the principal curvatures of the surface.

\begin{figure*}
\includegraphics[width=0.384\columnwidth]{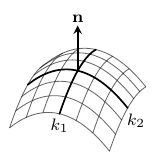} 
\hspace{12mm}
\includegraphics[width=0.557\columnwidth]{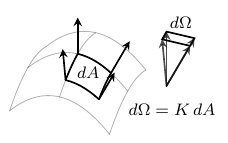} 
\hspace{3mm}
\includegraphics[width=0.557\columnwidth]{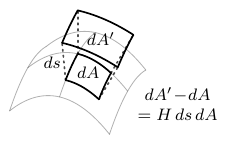}
\caption{Illustrating the properties of a curved surface. Left: The principal curvatures $k_1$ and $k_2$ are the curvatures of curves along the surface in the principal directions. Center: The solid angle $d\Omega$ swept out by the normal vectors on a small patch of area $dA$ is proportional to the Gaussian curvature $K$. Right: When a patch is slightly pushed out along the normal direction, the change in its area is proportional to the mean curvature $H$.}
\label{fig:surfaces}
\end{figure*}

Particularly symmetric surfaces can have the same principal curvatures at every point. For example, a plane always has $k_1 = k_2 = 0$, a sphere of radius $R$ has $k_1 = k_2 = 1/R$, and a cylinder of radius $R$ has one principal curvature equal to zero and the other equal to $1/R$. Note that the ordering of $k_1$ and $k_2$ is arbitrary, and their signs depend on the choice of sign for the normal vector, so only the sign of their product is meaningful. 

Given the principal curvatures, it is useful to define the Gaussian curvature $K$ and mean curvature $H$ by 
\begin{align}
K &= k_1 k_2, \\
H &= k_1 + k_2.
\end{align}
They have simple geometric meanings in terms of the surface's normal vector, which is
\begin{equation} \label{eq:n_vec}
\n \simeq (k_1 x, k_2 y, 1 - (k_1 x)^2/2 - (k_2 y)^2/2),
\end{equation}
normalized up to cubic and higher-order terms. 

Consider an infinitesimal patch of the surface, bounded by $0 \leq x, y \leq d\ell$. The patch has area $dA = (d\ell)^2$, and the unit normal vectors on the patch sweep out a solid angle $d\Omega=(k_1 \, d\ell) (k_2 \, d\ell)$. Thus, in general the area of a patch is related to the solid angle $d\Omega$ swept out by its normal vectors by 
\begin{equation} \label{eq:area_elt}
K \, dA = d\Omega,
\end{equation}
as shown in Fig.~\ref{fig:surfaces}. As for the mean curvature, taking the divergence of the normal vector at the origin yields
\begin{equation} \label{eq:div_n}
\grad \cdot \n = H.
\end{equation}
Technically, to take the divergence of the normal vector, we must extend it to also be defined off the surface. However, Eq.~\eqref{eq:div_n} is valid because the divergence has the same value regardless of the choice of extension.  

These two results show the qualitative difference between the Gaussian and mean curvatures. Consider deforming a surface in space, without locally stretching it. Such deformations do not change the area $dA$ of a small patch by definition, and it can be shown that they also do not change $d\Omega$. (For a visual proof, see Ref.~\cite{needham2021visual}.) Thus, they do not affect the Gaussian curvature, which therefore describes the surface's ``intrinsic'' curvature. By contrast, the mean curvature is ``extrinsic'': it depends on how the surface is embedded in three-dimensional space. For instance, a flat piece of paper can be rolled into a cylinder, which has nonzero mean curvature. 

A remarkable result follows from integrating Eq.~\eqref{eq:area_elt} over a closed orientable surface $S$,
\begin{equation}
\int_S K \, dA = \int_S d\Omega.
\end{equation}
The left-hand side is called the total curvature. For a sphere, the right-hand side is $4\pi$ since every possible normal direction $\n$ occurs at precisely one point on the surface. This remains true if the sphere is arbitrarily deformed while remaining convex. As illustrated in Ref.~\cite{needham2021visual}, more dramatic deformations can produce multiple points with the same $\n$. However, the additional contributions to the oriented solid angle integral cancel in pairs of points with $\n$ of the same direction but opposite sign, leaving the integral the same. These considerations lead to the Gauss--Bonnet theorem: the total curvature only depends on the surface's topology. It is $4\pi$ for a sphere, zero for a torus since the parts near and away from the hole contribute $\pm 4\pi$, and more generally can be determined in terms of the surface's number of holes. 

As for Eq.~\eqref{eq:div_n}, consider pushing a patch $dA$ of the surface along the normal vector for a distance $ds$, as shown in Fig.~\ref{fig:surfaces}. Integrating $\grad \cdot \n$ in the volume swept out between these patches, using the divergence theorem, yields 
\begin{equation} \label{eq:area_deform}
H \, ds \, dA = dA' - dA.
\end{equation}
Thus, the mean curvature determines how the surface's area changes when it is expanded,
\begin{equation}
\frac{dA}{ds} = H A.
\end{equation}

For example, consider a physical surface with surface tension $\gamma$ and a pressure difference $\Delta P$ across it. In equilibrium, the work done by the pressure under a virtual displacement of the surface balances the change in surface energy, $(\Delta P) A \, ds = \gamma\, dA = \gamma H A \, ds$, which implies the Young--Laplace equation $\Delta P = \gamma H$. In the case of a soap film bounded by a frame, the pressure on both sides is atmospheric, so $\Delta P$ is zero. In equilibrium, such a film forms an open surface with zero mean curvature, and it also locally minimizes its area. Accordingly, surfaces with zero mean curvature are called minimal surfaces. 

Finally, we will need to know how the principal radii of curvature $R_i$ vary when the surface is pushed out along the normal vector by a distance $ds$. Intuitively, a curve's radius of curvature is simply the distance to its center of curvature, so $dR_1 = dR_2 = ds$. 

We can also show this explicitly. For each point $\r=(x,y,z(x,y))$ on the original surface, we can construct the shifted point $\r' = \r + \n \, ds = (x',y',z'(x',y'))$, where $\n$ is given by Eq.~\eqref{eq:n_vec}. This gives $x'\simeq(1+k_1\,ds) \, x$ and $y'\simeq(1+k_2\,ds) \, y$, and using Eq.~\eqref{eq:z_param} for $z(x, y)$ also
\begin{equation}
z'\simeq ds-\frac12k_1x^2(1+k_1\,ds)-\frac12k_2y^2(1+k_2\,ds).
\end{equation}
Upon eliminating $x$ and $y$ in favor of $x'$ and $y'$, we get
\begin{equation}
-z'+ds \simeq \frac{k_1'}{2} \, x'^2 + \frac{k_2'}{2} \, y'^2,
\end{equation}
where $k_i' = k_i / (1 + k_i \, ds)$. This has the same structure as Eq.~\eqref{eq:z_param} except for an overall shift by $ds$ along the $z$-axis. We find that the principal radii of the new surface are $R_i'=1/k_i'=R_i+ds$, as anticipated.


\section{Electrostatic Fields With Straight Field Lines}
\label{sec:straightlines}

We now return to the classification of electrostatic fields with straight field lines. If the field lines are straight, then the unit vector $\n$ describing the field's direction does not change when we move along $\n$, so 
\begin{equation}
(\n \cdot \grad) \, \n = \ve{0}.
\end{equation}
This implies that the curl of $\n$ is parallel to $\n$, as
\begin{equation} \label{eq:curl_condition}
\n \times (\grad \times \n) = \frac12 \, \grad(\n\cdot\n) - (\n \cdot \grad) \, \n = \ve{0}. 
\end{equation}
However, we showed in Sec.~\ref{sec:integrability} that $\n$ can correspond to a conservative electric field $\E = f \n$ only if the curl of $\n$ is orthogonal to $\n$. Since the curl of $\n$ is both parallel and orthogonal to $\n$, we must have
\begin{equation} \label{eq:no_curl}
\grad \times \n = \bm{0}.
\end{equation}
Thus, there is a function $u$ such that, locally, $\n = \grad u$.

\begin{figure}
\includegraphics[width=0.5\columnwidth]{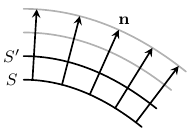}
\caption{If a conservative field has straight field lines, its equipotentials must be parallel surfaces: given an equipotential surface $S$, another equipotential surface $S'$ is reached by following the normal direction by the same distance starting from any point on $S$.}
\label{fig:parallel}
\end{figure}

The quantity $u$, labeling equipotential surfaces of $\n$, has a simple interpretation. Take two equipotential surfaces $S$ and $S'$ and express the difference of their values of $u$ in terms of a line integral along the field line of $\n$,
\begin{equation}
u_{S'}-u_S=\int\grad u\cdot d\r=\int\grad u\cdot\n\,ds=\int ds,
\end{equation}
where $ds=|d\r|$ is the length of an infinitesimal line element, and we used the fact that $\n$ is a unit vector. We see that the change in $u$ measures the distance between equipotential surfaces along the normal direction. As a consequence, any two equipotential surfaces are parallel: as illustrated in Fig.~\ref{fig:parallel}, the distance between them is the same at all points.

Since the curl of $\n$ is zero, the curl of $\E = f \n$ is equal to $(\grad f) \times \n$. Thus, the field is conservative if its magnitude $f$ is constant on equipotential surfaces. Equivalently, $f$ can be expressed in terms of $u$, so that
\begin{equation}
\E(\r) = f(u(\r)) \, \n(\r)
\end{equation}
for an arbitrary function $f(u)$. 

This implies that there are many electrostatic fields with straight field lines, beyond the textbook examples with spherical, cylindrical or planar symmetry. Concretely, given a single equipotential surface $S$ of arbitrary shape, we can construct such a field by defining other equipotential surfaces as parallel to $S$, and picking an arbitrary field magnitude $f(u)$ on each equipotential surface. The construction can be extended until the field becomes singular: e.g., when the field lines collide. If $S$ is closed and convex, this will never occur outside of $S$, so a conservative field with straight field lines can be constructed everywhere outside of it. However, in the context of electrostatics, the field may have a complicated charge density profile, and we do not know of nontrivial examples with simple physical interpretations. 

We now consider the more restrictive case of vacuum electrostatic fields, which additionally have zero divergence. In general, the divergence is 
\begin{equation}
\grad \cdot \E = (\n \cdot \grad) f + (\grad \cdot \n) \, f.
\end{equation}
To simplify the first term, note that it is $\n \cdot (\grad f) = \n \cdot \grad u \, (df/du) = df/du$, where we evaluated $\grad f$ by the chain rule. As for the second term, apply Eq.~\eqref{eq:div_n} to write it in terms of the mean curvature $H$ of the equipotential surface at that point. The result~is 
\begin{equation}
\grad \cdot \E = \frac{df}{du} + H f. \label{eq:div_formula}
\end{equation}
Setting the divergence to zero yields the constraint
\begin{equation} 
H = - \frac{1}{f} \frac{df}{du}. \label{eq:zero_div_eq}
\end{equation}
Crucially, since $f$ is a function of $u$, the right-hand side takes the same value at every point on an equipotential surface. Therefore, each equipotential surface must have constant mean curvature. 

As a simple example, the electric field of a point charge has spherical equipotential surfaces, and a sphere indeed has constant mean curvature. We may take $u$ to be equal to the sphere's radius $r$, in which case $H = 2/u$. Imposing Eq.~\eqref{eq:zero_div_eq} yields the inverse square law $f \propto 1/u^2$. 

There are many nontrivial surfaces with constant mean curvature, but a stronger constraint arises from considering two nearby equipotential surfaces. Take a point on an equipotential surface $S$ with principal radii of curvature $R_1$ and $R_2$, and consider moving along the normal by a distance $ds$ to a new point, on another equipotential surface $S'$. As discussed at the end of Sec.~\ref{sec:curvature}, the radii of curvature of $S'$ at that new point are $R_1 + ds$ and $R_2 + ds$, so the mean curvature is
\begin{equation}
H' = \frac{1}{R_1 + ds} + \frac{1}{R_2 + ds} = H + (2 K - H^2) \, ds
\end{equation}
to first order in $ds$. But $S'$ must also have constant mean curvature, so the Gaussian curvature $K$ must also have been constant on $S$. Since $H$ and $K$ determine the principal curvatures $k_1$ and $k_2$, each equipotential surface must have constant principal curvatures. 

We have thus reduced our problem of finding vacuum electrostatic fields with straight field lines to a purely geometrical question: what is the most general family of parallel surfaces with constant principal curvatures? To begin, we parametrize a single surface near the origin as $\r(x, y) = (x, y, z(x, y))$, in terms of the two coordinates $(x_1, x_2) = (x, y)$. (This notation differs from Sec.~\ref{sec:integrability}, where $\r$ stood for an arbitrary point in space.) 

Next, we define some quantities to extract the principal curvatures. The vectors describing motion along the surface, per unit change in the coordinates, are $\ve{e}_i = \del \r / \del x_i$. Explicitly, if we define $z(x,y)$ by Eq.~\eqref{eq:z_param}, they are
\begin{align}
\ve{e}_x &= \frac{\del \r}{\del x} \simeq (1, 0, -k_1 x), \label{eq:ex_def} \\
\ve{e}_y &= \frac{\del \r}{\del y} \simeq (0, 1, -k_2 y). \label{eq:ey_def}
\end{align}
Now, as discussed in Sec.~\ref{sec:curvature}, curvature describes how the normal vector $\n$ varies along the surface. Part of this information is captured in the ``second fundamental form'' $\ve{B}$, a $2 \times 2$ matrix with elements 
\begin{equation} \label{eq:B_def}
B_{ij} = \ve{e}_i \cdot \frac{\del \n}{\del x_j}.
\end{equation}
At the origin, the eigenvalues of $\ve{B}$ are the principal curvatures, but this is not true elsewhere because distances along the surface in the $(x, y)$ coordinate system become distorted. To account for this, we define the metric or ``first fundamental form'' $\ve{G}$, which is another $2 \times 2$ matrix with elements 
\begin{equation} \label{eq:G_def}
G_{ij} = \ve{e}_i \cdot \ve{e}_j.
\end{equation}
This is equal to the identity matrix at the origin, but differs elsewhere. However, under a change of coordinates, $B_{ij}$ and $G_{ij}$ pick up the same Jacobian factors. Thus, the eigenvalues of the matrix $\ve{G}^{-1} \ve{B}$ are invariant under coordinate changes, and are equal to the principal curvatures at every point on the surface. 

For a reader wishing to learn more about the geometry of surfaces, the first and second fundamental forms are introduced, with the component notation we use here, in Eqs.~(8.1) and (8.7) of Ref.~\cite{Frankel2012a}, and below definition 3.10 of Ref.~\cite{kuhnel2015differential}. Note that our definition of $\ve{B}$ differs from these references by a sign for convenience.

For surfaces parametrized by the Cartesian coordinates $x$ and $y$, an alternative form for $\ve{B}$ is more useful than the definition~\eqref{eq:B_def}. Since $\n \cdot \ve{e}_i$ is zero, we can use the product rule to write
\begin{equation} \label{eq:B_expr}
B_{ij} = -\frac{\del \ve{e}_i}{\del x_j} \cdot \n = -\frac{\del^2 \r}{\del x_i \del x_j} \cdot \n = \frac{\del^2 (-z)}{\del x_i \del x_j} \, n_z,
\end{equation}
where the final step follows because the second partial derivatives of $x$ and $y$ vanish. Thus, we can evaluate $\ve{B}$ knowing only $z$ and $n_z$, without having to calculate any intermediate quantities. 

We are now prepared to evaluate the principal curvatures of a surface near the origin. To warm up, consider the surface defined by Eq.~\eqref{eq:z_param}, where $z$ is quadratic in $x$ and $y$. Using Eq.~\eqref{eq:B_expr}, we immediately read off 
\begin{equation}
\ve{B} \simeq \begin{pmatrix} k_1 & 0 \\ 0 & k_2 \end{pmatrix} n_z
\end{equation}
where $n_z$ is given by Eq.~\eqref{eq:n_vec}. We calculate $\ve{G}$ using Eqs.~\eqref{eq:ex_def} and~\eqref{eq:ey_def}, and inverting it gives, to second order,  
\begin{equation}
\ve{G}^{-1} \simeq \begin{pmatrix} 1 - (k_1 x)^2 & -k_1 k_2 x y \\ -k_1 k_2 x y & 1 - (k_2 y)^2 \end{pmatrix}.
\end{equation}
The eigenvalues of $\ve{G}^{-1} \ve{B}$ are \textit{not} independent of $x$ and $y$. For instance, when $y = 0$ we have, to second order,
\begin{equation} \label{eq:curvature_quadric}
\ve{G}^{-1} \ve{B} \simeq \begin{pmatrix} k_1 (1 - 3 k_1^2 x^2/2) & 0 \\ 0 & k_2 (1 - k_1^2 x^2/2) \end{pmatrix} 
\end{equation}
whose eigenvalues have $(k_1 x)^2$ terms. Similarly, when $x = 0$ the eigenvalues have $(k_2 y)^2$ terms. It thus naively appears that, besides the trivial case $k_1 = k_2 = 0$, a surface cannot have constant principal curvatures.

Of course, there do exist surfaces with constant nonzero principal curvatures, such as the sphere. The subtlety is that, to evaluate these curvatures to second order in $x$ and $y$, we need to evaluate both $\ve{B}$ and $\ve{G}$ to second order. However, $\ve{B}$ involves a second derivative of $z$, so correctly computing it to second order requires expanding $z$ to \textit{fourth} order. To cancel off the curvatures' $x^2$ and $y^2$ terms, the fourth order terms must be even in $x$ and $y$. Thus, it suffices to take
\begin{equation} \label{eq:z_taylor_full}
-z = \frac{k_1 x^2 + k_2 y^2}{2} + \frac{c_1 x^4 + c_2 y^4}{24} + \frac{c_3 x^2 y^2}{4},
\end{equation}
where the numeric coefficients are chosen for later convenience. Now, to second order, $\ve{B}$ becomes
\begin{equation}
\ve{B}\! \simeq \! \begin{pmatrix} k_1 \!+\! (c_1 x^2 + c_3 y^2)/2 \! \! & c_3 x y \\ c_3 x y & \! \! k_2 \!+\! (c_2 y^2 + c_3 x^2) / 2 \end{pmatrix}\! n_z.
\end{equation}
These additional terms qualitatively change the situation. For example, when $y = 0$ we have, to second order,
\begin{equation}
\ve{G}^{-1} \ve{B} \! \simeq \! \begin{pmatrix} k_1 \!+\! (c_1 \!-\! 3 k_1^3) x^2/2 \! \! & 0 \\ 0 & \! \! k_2 \!+\! (c_3 \!-\! k_1^2 k_2) x^2/2 \end{pmatrix}
\end{equation}
where the second order terms cancel if $c_1 = 3 k_1^3$ and $c_3 = k_1^2 k_2$. Similarly, considering $x = 0$ yields the requirements $c_2 = 3 k_2^3$ and $c_3 = k_1 k_2^2$. 

Crucially, we find four constraints on only three independent parameters. The two constraints on $c_3$ can only be simultaneously satisfied if $k_1^2 k_2 = k_1 k_2^2$, which implies
\begin{equation} \label{eq:k_condition}
k_1 k_2 (k_1 - k_2) = 0.
\end{equation}
In other words, for the principal curvatures to be constant, even locally, at least one of the $k_i$ must be zero, yielding a plane or cylinder, or they must be equal, yielding a sphere. Perhaps surprisingly, it is impossible to have a surface with constant principal curvatures where $k_1,k_2$ are different and both nonzero. 

We conclude that the only vacuum electrostatic fields with straight field lines are the familiar examples generated by a point charge, line charge, or plane of charge. 


\section{Conclusions}
\label{sec:discussion}

We have shown that a set of electric field lines can correspond to an electrostatic field if it obeys the integrability condition~\eqref{eq:integratingfactor}, and that vacuum electrostatic fields with straight field lines must have planar, cylindrical, or spherical symmetry. Along the way, we have given a brief introduction to some fundamental geometric ideas. 

In the spirit of Feynman's remark that ``the same equations have the same solutions'', we note that our results apply to a variety of other physical contexts. For example, Eq.~\eqref{eq:integratingfactor} can also be used to determine when a given family of field lines corresponds to a magnetostatic field in a current-free region, to a gravitational field, or to the streamlines of an irrotational fluid flow. 

Vacuum electrostatic fields with straight field lines have been discussed elsewhere. In Ref.~\cite{Urbantke1994a}, the same result is derived with a subtle heuristic geometric argument involving the caustic surfaces formed by the field lines, and by an elaborate integrability argument involving the second and third derivatives of $u$. In the mathematical literature, Ref.~\cite{Martinez2000a} gives a more formal argument using isothermal coordinates and differential forms. Finally, the condition~\eqref{eq:k_condition} is equivalent to theorem 7.55 of Ref.~\cite{Montiel2009}. However, our argument presupposes only vector calculus, and may be the most direct and elementary. 

If nonzero charge density is allowed, we have found a prescription to generate many electrostatic fields with straight field lines: the divergence of the field is simply given by Eq.~\eqref{eq:div_formula}. This construction works for any initial equipotential surface $S$ and any choice of field magnitude $f(u)$. The result is an electrostatic field with constant magnitude on every equipotential, which allows the Gaussian flux integral to be easily evaluated even in the absence of spherical, cylindrical, or planar symmetry. Developing this idea further may produce useful examples for electromagnetism courses. The same construction applies for gravitational fields, which have the additional constraint of non-negative divergence. Referring to Eq.~\eqref{eq:div_formula}, this constraint can always be satisfied by choosing $f(u)$ to have a sufficiently large derivative.

For further discussion, we direct the interested reader to Ref.~\cite{franklin2024taxonomy}, which considers the topology of magnetostatic field lines, and section 12.4 of Ref.~\cite{needham2023complex}. In our language, the latter discusses when equipotentials in \textit{two} dimensions can correspond to divergence-free fields. However, the geometric tools used in that work are specific to two dimensions, and we have not found a simple answer for the analogous question in three dimensions. 

\begin{acknowledgments}
T.B.~thanks Gard Zakarias Stadheim for raising the question of electrostatic fields with straight field lines. We are indebted to Pavel Gumenyuk for collaboration in the early stages of this project and for useful feedback on the manuscript. Last but not least, we thank the \href{https://physics.stackexchange.com/questions/610992/electrostatic-fields-with-straight-field-lines}{Physics StackExchange} community for discussions.
\end{acknowledgments}

\section*{Author declaration}

The authors have no conflicts of interest to disclose.

\bibliography{refs}

\end{document}